\documentclass[reprint,amsmath,amssymb,aps,pra]{revtex4-2}
\usepackage{mathrsfs}
\usepackage{color}
\usepackage{graphicx}
\usepackage{dcolumn}
\usepackage{bm}
\usepackage[normalem]{ulem}
\usepackage{hyperref}
\usepackage{hypernat}
\usepackage{pstricks}
\usepackage{amsmath,amssymb}
\usepackage{version}
\usepackage{ulem}
\usepackage{mathrsfs}
\usepackage{lipsum}
\usepackage{siunitx}
\usepackage{braket}
\usepackage[justification=raggedright, singlelinecheck=false]{caption}

\begin{document}

\preprint{APS/123-QED}

\title{Coherent Control of Ion-Photoelectron Dynamics through Rabi Oscillations: An \textit{ab initio} study}
\author{Bo-Ren Shen$^{1}$}
\thanks{These authors contributed equally to this work.}

\author{Yi-Jia Mao$^{1,2}$}
\thanks{These authors contributed equally to this work.}
\author{Zhao-Han Zhang$^{1}$}
\author{Yang Li$^{1}$}\email{liyang22@sjtu.edu.cn}
\author{Takeshi Sato$^{3,4,5}$}
\author{Kenichi L. Ishikawa$^{3,4,5,6}$}\email{ishiken@n.t.u-tokyo.ac.jp}
\author{Feng He$^{1,2}$}\email{fhe@sjtu.edu.cn}
\affiliation{
    $^1$Key Laboratory for Laser Plasmas (Ministry of Education) and School of Physics and Astronomy, Collaborative Innovation Center for IFSA (CICIFSA), Shanghai Jiao Tong University, Shanghai 200240, China\\
    $^2$Tsung-Dao Lee Institute, Shanghai Jiao Tong University, Shanghai 201210, China\\
    $^3$Department of Nuclear Engineering and Management, Graduate School of Engineering, The University of Tokyo,7-3-1 Hongo, Bunkyo-ku, Tokyo 113-8656, Japan\\
    $^4$Photon Science Center, Graduate School of Engineering, The University of Tokyo, 7-3-1 Hongo, Bunkyo-ku, Tokyo 113-8656, Japan\\
    $^5$Research Institute for Photon Science and Laser Technology, The University of Tokyo, 7-3-1 Hongo, Bunkyo-ku, Tokyo 113-0033, Japan\\
    $^6$Institute for Attosecond Laser Facility, The University of Tokyo, 7-3-1 Hongo, Bunkyo-ku, Tokyo 113-0033, Japan
}

\date{ \today}
	
\begin{abstract}
We present first-principles numerical simulations of photoionization in neon induced by bichromatic extreme ultraviolet pulses with frequencies $\omega$ and $2\omega$, specially chosen to make $\omega$ equal to the energy difference between the $2s$ and $2p$ subshells. This allows for the production of photoelectrons from the $2s$ shell by $2\omega$ pulse and from the $2p$ shell by $\omega$ pulse with the same energy. Using the multi-configurational time-dependent Hartree-Fock method, we explore how Rabi coupling between subshells generates coherence between the corresponding photoelectron wave packets. Our \textit{ab initio} calculations confirm the analytical results derived from the essential-states approach in [K. L. Ishikawa, K. C. Prince, and K. Ueda, J. Phys. Chem. A 127, 10638 (2023)], validating the theoretical predictions. Although we focus on the Ne $2p$ and $2s$ subshells, our approach is applicable to a broad range of systems exhibiting photoionization from multiple subshells. The laser parameters employed in our simulations are available in modern Free Electron Lasers (FELs), and we anticipate that this work could stimulate experimental investigations using FELs to study ion-photoelectron coherence and entanglement.
\end{abstract}

\maketitle

\section{Introduction}

Coherent control leverages the ability to manipulate the dynamics of quantum systems using independently controllable parameters, enabling the study of rapid processes and demonstrating fundamental quantum principles across atomic physics, solid-state physics, chemistry, and biology~\cite{feist2015quantum,you2016attosecond,mahmood2016selective,zewail2000femtochemistry}. Central to this concept is the control over interference between multiple transition pathways, which can be achieved using bichromatic fields, such as those consisting of a fundamental frequency $\omega$ and its higher harmonics~\cite{yin1992asymmetric,prince2016coherent,giannessi2018coherent,di2019complete,you2020new,popova2022ionization,mao2023unveiling,chen2025resolving}. These fields allow for independent manipulation of parameters like phase, intensity, and polarization to control the ionization process, leading to phenomena such as interference in photoionization.
%This approach of combining the fundamental and higher harmonics can also be extended to the strong-field regime~\cite{yao2023multiphoton}.%
Coherent control in the optical range was first demonstrated in mercury atoms using $\omega$+$3\omega$ radiation~\cite{chen1990interference}. With the advancement of high-power lasers and FELs, this technique has been extended into the vacuum ultraviolet (VUV), extreme ultraviolet (XUV), and even X-ray ranges~\cite{ayvazyan2002generation,ackermann2007operation,emma2010first,allaria2012highly,young2018roadmap}. The FEL technology, with its ability to generate high-frequency coherent radiation, has enabled groundbreaking experiments in photoionization~\cite{prince2016coherent,giannessi2018coherent,di2019complete,you2020new,popova2022ionization,richter2024strong,nandi2022observation}.
%hui2023ultrafast,yang2023harmonic%
For instance, in photoionization experiments with neon, the use of $\omega$+$2\omega$ bichromatic light results in photoelectrons with the same energy but opposite parities, where the interference between ionization pathways modulates the photoelectron angular distribution (PAD) and its asymmetry as a function of the relative phase between the components~\cite{prince2016coherent}. Another example is a photoionization experiment on helium, where the PAD obtained through interference from multiple pathways can be used to infer the absolute phase relationship and amplitude ratio between the fundamental frequency and its second harmonic in a bichromatic XUV beam~\cite{di2019complete}. It is directly applied at the experimental station, allowing it to automatically include phase changes that may be introduced during beam transmission.

The coherent control of photoionization is typically applied to signals originating from the same shell. In contrast, for signals from different shells, even if the frequencies of the bichromatic fields are precisely tuned such that the photoelectrons from different shells share the same energy, they remain incoherent. This incoherence arises from the entanglement between the photoelectron and the ionic states, forming a composite biparticle system together~\cite{laurell2025measuring}. This ion-electron entanglement has gained growing interest in the photoionization of multi-electron systems in recent research~\cite{vatasescu2013entanglement,vrakking2021control,koll2022experimental,ruberti2024bell}. For example, the vibrational coherence of ${\rm H}_2$ has been investigated and controlled both theoretically and experimentally using two time-delayed attosecond pulses~\cite{vrakking2021control,koll2022experimental}. Moreover, a numerical Bell test has been proposed to demonstrate entanglement in the strong-field ionization of argon atoms~\cite{ruberti2024bell}.

Very recently, an inspiring scheme for XUV $\omega$+$2\omega$ bichromatic photoionization of a neon atom has been theoretically proposed by one of us~\cite{ishikawa2023control}, where the incoherent channels regain coherence due to Rabi oscillations between two sub-shells. In this scheme, the fundamental photon energy $\omega$ is tuned to the energy difference (26.91\,eV) between the ionization thresholds for Ne $2p$ (21.56\,eV) and $2s$ (48.47\,eV) states. The ionization processes from the $2s$ and $2p$ states result in photoelectrons with identical energies (5.35\,eV) but opposite parities by absorbing one $2\omega$ photon or $\omega$ photon, respectively. However, the photoelectron wave packets originating from the two ionization pathways do not exhibit interference. Interestingly, when the fundamental pulse couples the two ionic states ($2s^{-1}$ and $2p^{-1}$), inducing Rabi oscillations~\cite{rabi1937space,gentile1989experimental,windpassinger2008nondestructive,flogel2017rabi,yu2018core,nandi2022observation}, coherence can be created between the photoelectrons from the two pathways. By using an essential-states approach~\cite{yu2018core,haan1995effects,eberly1991above,grobe1993observation}, the authors show that this Rabi coupling converts the initial incoherent photoelectron wave packets into a coherent superposition, which can be controlled by parameters such as the relative phase, pulse width, delay between the wavelength components, Rabi coupling strength, and photoelectron energy. 

On the theoretical side, accurately capturing such dynamics of correlated electrons and ions calls for methods that are capable of describing correlated multielectron dynamics within the framework of a time-dependent, strong-field approach. Such methods provide a rigorous numerical benchmark for simplified analytical models, which approximately include some description of multielectron interactions but not the full quantum picture. Among them, the multiconfigurational time-dependent Hartree-Fock (MCTDHF) method and its extensions are good candidates~\cite{ishikawa2015review,hochstuhl2014time,lode2020colloquium,zanghellini2003mctdhf,kato2004time}, which permit a good description of multielectron effects and apply to strong field regimes~\cite{sato2013time,miyagi2013time,sato2015time}. 

In this follow-up to previous work \cite{ishikawa2023control}, we present first-principle numerical simulations of photoionization by a pair of coherent, ultrashort, fundamental, and second-harmonic XUV pulses, designed to ionize two different subshells with the same photoelectron energy. Using MCTDHF, we investigate how the Rabi coupling between subshells, induced by the resonant photon of the fundamental pulse, generates coherence between the corresponding photoelectron wave packets. The coherence, restored by Rabi coupling, manifests as an asymmetry in the energy-resolved PAD, governed by the relative phase of the bichromatic pulses. Interference fringes also appear in the total photoelectron momentum distribution, which is otherwise absent without Rabi coupling. Our MCTDHF calculations confirm the analytical results obtained in the previous work using the essential-states approach, providing a full-dimensional all-electron description of the ionization dynamics. This study enhances the understanding of coherent control of multi-photon ionization dynamics and provides a comprehensive numerical framework for exploring ion-photoelectron entanglement and coherence.

The structure of this paper is as follows. Section~\ref{secII} outlines the numerical methods employed in this study. In Sec.~\ref{secIII}, the effect of different relative phases on the photoelectron spectrum is discussed. Additionally, by reducing the time delay and integrating the asymmetry over a specific energy range, the feasibility of observing asymmetry in low-energy resolution experiments is demonstrated. Section~\ref{secIV} concludes with a summary of the paper.

\section{Numerical Methods \label{secII}}
The dynamics of a multi-electron atom in intense laser fields is described by the time-dependent Hamiltonian
\begin{equation}
    \hat{H}(t) = \sum_{i=1}^N \hat{h}_1(\mathbf{r}_i,t) + \sum_{i<j}^N \frac{1}{|\mathbf{r}_i - \mathbf{r}_j|},
    \label{eq:Hamiltonian}
\end{equation}
where $N$ is the number of electrons. The single-particle Hamiltonian in the length gauge is given by
\begin{equation}
    \hat{h}_1(\mathbf{r},t) = -\frac{1}{2}\nabla^2 - \frac{Z}{r} + \mathbf{E}(t)\cdot\mathbf{r},
    \label{eq:h1_length}
\end{equation}
with the laser electric field
\begin{align}
    \mathbf{E}(t) &= E_\omega \sin^2\left(\frac{\omega t}{2N_1}\right)\cos(\omega t)\mathbf{e} _z \nonumber \\
    &\quad + E_{2\omega} \sin^2\left[\frac{2\omega(t-\tau)}{2N_2}\right]\cos\left[2\omega(t-\tau)- \delta\right]\mathbf{e} _z,
    \label{eq:Efield}
\end{align}
where $E_\omega$ and $E_{2\omega}$ are field amplitudes, $\tau$ is the time delay between pulses, and $\delta$ is the relative phase. 

In the MCTDHF method, the total wavefunction is expanded in a basis of time-dependent Slater determinants
\begin{equation}
    \Psi(t) = \sum_I C_I(t) \Phi_I(\{\psi_p\},t),
    \label{eq:psi_expansion}
\end{equation}
where $C_I(t)$ are configuration interaction (CI) coefficients and $\Phi_I$ are determinants constructed from orthonormal single-particle orbitals $\psi_p(\mathbf{r},t)$. 

Applying the time-dependent variational principle to the action functional
\begin{equation}
    S = \int_{t_0}^{t_1} \bra{\Psi} \hat{H} - i\partial_t \ket{\Psi} dt,
    \label{eq:action}
\end{equation}
we derive coupled equations of motion for the CI coefficients and orbitals. The CI coefficients evolve according to
\begin{equation}
    i\partial_t C_I = \sum_J \bra{\Phi_I} \hat{H} - i\hat{X} \ket{\Phi_J} C_J,
    \label{eq:EOMc}
\end{equation}
where the matrix elements of the operator $\hat{X}$ are defined as $X^p_q = \bra{\psi_p}\partial_t\ket{\psi_q}$. The orbital dynamics are governed by
\begin{equation}
    i\partial_t \ket{\psi_p} = \hat{Q} \left( \hat{h}_1 + \hat{F} \right) \ket{\psi_p},
    \label{eq:EOMo}
\end{equation}
with the projection operator $\hat{Q} = 1 - \sum_{q=1}^M \ket{\psi_q}\bra{\psi_q}$ ensuring orthogonality, where $M$ is the number of occupied orbitals. The nonlocal operator $\hat{F}$ contains electron correlation effects
\begin{equation}
    \hat{F}\ket{\psi_p} = \sum_{o,q,s,r}^{\text{occ}} (D^{-1})^o_p P^{qs}_{or} \hat{W}^r_s \ket{\psi_q},
    \label{eq:F_operator}
\end{equation}
where the mean-field potential is
\begin{equation}
    \hat{W}^r_s(\mathbf{r}) = \int \frac{\psi_r^*(\mathbf{r}')\psi_s(\mathbf{r}')}{|\mathbf{r}-\mathbf{r}'|} d\mathbf{r}',
    \label{eq:meanfield}
\end{equation}
and $D$ and $P$ are the one-body and two-body reduced density matrices, respectively.

Our numerical implementation employs a spherical finite-element discrete variable representation (FE-DVR) for the single electron orbitals. In particular, we perform MCTDHF simulations in which each orbital is expanded in spherical harmonics up to a maximum angular momentum quantum number of $l=12$. The radial domain extends to 120 a.u. and is partitioned into 30 finite elements, each containing 23 DVR grid points. The time propagation employs a time step of $1/1000$ of the optical cycle corresponding to the $\omega$ pulse. The initial ground state is obtained by imaginary time propagation of the equations of motion. We use 5 frozen-core orbitals and 9 active orbitals, constituting the complete active space to accommodate 8 electrons in all the calculations.

To analyze the influence of Rabi dynamics on photoelectron coherence, it is advantageous to obtain ion-channel-resolved photoelectron spectra. However, in the MCTDHF framework, the ionic states are predefined, making it challenging to identify the specific ion state with which the photoelectron is entangled. To address this, we have developed a channel-resolved time-dependent surface flux (t-SURFF) method that enables the resolution of the entangled electron-ion state and the extraction of channel-resolved photoelectron spectra. Further details will be presented elsewhere.

\section{Results and Discussions \label{secIII}}
\begin{figure*}[htbp]
\centering
\includegraphics[width=\textwidth]{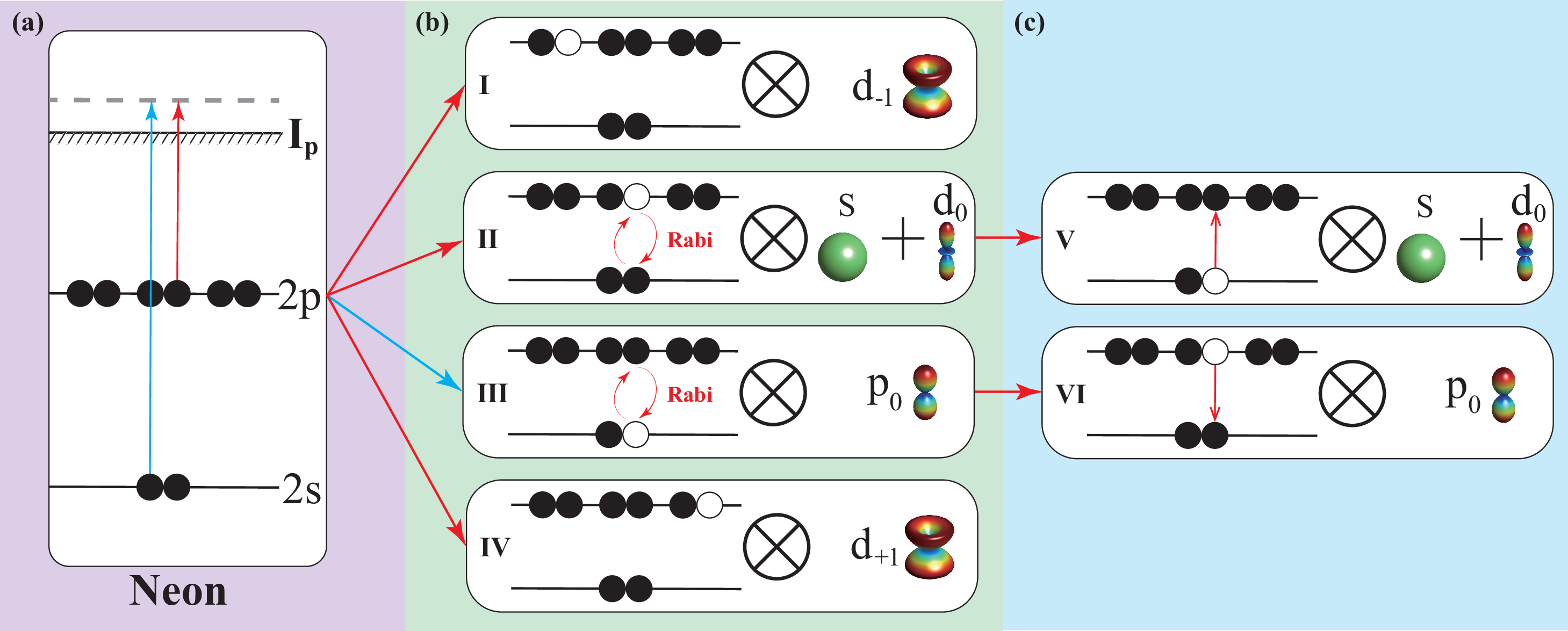}
\captionsetup{width=\textwidth}
\caption{(a) Schematic diagrams illustrating the photoionization of neon atoms by a $\omega+2\omega$ pulse pair. (b) The single-photon processes: \uppercase\expandafter{\romannumeral1}. Ionization of the $2p_{-1}$ orbital by a photon with energy $\omega$, where $\omega$ is the resonant energy for the $2s$ to $2p$ transition, results in the ionic state $2p_{-1}^{-1}$ and a $d_{-1}$ state photoelectron; \uppercase\expandafter{\romannumeral2}. Ionization of the $2p_{0}$ orbital by a photon with energy $\omega$ results in the ionic state $2p_{0}^{-1}$, with $s$ state electron and $d_{0}$ state electron; \uppercase\expandafter{\romannumeral3}. Ionization of the $2s$ orbital by a photon with energy $2\omega$ results in the ionic state $2s^{-1}$ and $p_{0}$ state electron; \uppercase\expandafter{\romannumeral4}. Ionization of the $2p_{+1}$ orbital by a photon with energy $\omega$ results in the ionic state $2p_{+1}^{-1}$ and a $d_{+1}$ state photoelectron. (c) The two-photon processes: \uppercase\expandafter{\romannumeral5}. After process \uppercase\expandafter{\romannumeral2}, due to Rabi oscillations between the $2s$ and $2p$ states, an electron from the orbital $2p$ absorbs a photon with energy $\omega$ and transitions to the hole in the orbital $2s$; \uppercase\expandafter{\romannumeral6}. After process \uppercase\expandafter{\romannumeral3}, due to Rabi oscillations between the $2s$ and $2p$ states, an electron from the orbital $2p$ emits a photon with energy $\omega$ and transitions to the hole in the orbital $2p$.}
\label{diagram}
\end{figure*}

We investigate the photoionization of a neon atom induced by an $\omega+2\omega$ pulse pair, where $\omega$ is close to the energy difference between the ionization thresholds for the $2s$ and $2p$ orbitals in neon. Figure~\ref{diagram}(a) provides a schematic diagram of this photoionization process. When only single-photon processes are considered, five channels contribute to the photoelectron signal near $E_0=\omega-I_p$, where $I_p$ is the ionization threshold of the $2p$ orbital. The $2p$ electrons can scatter into $s_0$, $d_0$, or $d_{\pm 1}$ partial waves after absorbing an $\omega$ photon, generating the channels $\ket{2p_0^{-1}}\ket{s_0}$, $\ket{2p_0^{-1}}\ket{d_0}$, and $\ket{2p_{\pm1}^{-1}}\ket{d_{\pm1}}$, while the $2\omega$ pulse ionizes the atom from the $2s$ orbital, resulting in the $\ket{2s_0^{-1}}\ket{p_0}$ channel. The states $\ket{2s_0^{-1}}$ and $\ket{2p_{0,\pm1}^{-1}}$ denote ionic states with a hole in the $2s_0$ or $2p_{0,\pm1}$ orbitals, and $\ket{s_0}$, $\ket{p_0}$, and $\ket{d_{0,\pm1}}$ represent the partial waves of the photoelectron.

Although these five channels overlap in the photoelectron energy domain, only the contributions of $\ket{2p_0^{-1}}\ket{s_0}$ and $\ket{2p_0^{-1}}\ket{d_0}$ can interfere with each other. Therefore, interference between different photoelectron partial waves is only possible in identical ionic states. For simplicity, we combine these two channels into $\ket{2p_0^{-1}}\ket{s_0+d_0}$. The channels $\ket{2p_{-1}^{-1}}\ket{d_{-1}}$, $\ket{2p_0^{-1}}\ket{s_0+d_0}$, $\ket{2s_0^{-1}}\ket{p_0}$, and $\ket{2p_{+1}^{-1}}\ket{d_{+1}}$ correspond to pathways \uppercase\expandafter{\romannumeral1}, \uppercase\expandafter{\romannumeral2}, \uppercase\expandafter{\romannumeral3}, and \uppercase\expandafter{\romannumeral4}, respectively, as shown in Fig.~\ref{diagram}(b). Up to this point, the combination of these pathways only results in a symmetric angle- and energy-resolved photoelectron spectrum. This is because different pathways are incoherently summed due to their distinct ion states. Moreover, for pathway \uppercase\expandafter{\romannumeral2} (that is, channel $\ket{2p_0^{-1}}\ket{s_0+d_0}$), the partial waves of the photoelectron $\ket{s_0}$ and $\ket{d_0}$ have the same parity.

When considering two-photon processes, more channels are involved. If the $\omega$ pulse is long enough to induce Rabi oscillations between the $2s$ and $2p$ orbitals, the ion states $\ket{2s_{0}^{-1}}$ and $\ket{2p_{0}^{-1}}$ can be converted into each other. For the case shown in Fig.~\ref{diagram}(b) for pathway \uppercase\expandafter{\romannumeral2}, the electron in the $2s$ orbital can be excited to the $2p$ orbital by absorbing an additional $\omega$ photon, resulting in the $\ket{2s_0^{-1}}\ket{s_0+d_0}$ state, as depicted in Fig.~\ref{diagram}(c) for pathway \uppercase\expandafter{\romannumeral5}. Similarly, the electron in the $2p_0$ state of the $\ket{2s_0^{-1}}\ket{p_0}$ channel can transition to the $2s_0$ state, generating an additional channel, $\ket{2p_0^{-1}}\ket{p_0}$, as shown in Fig.~\ref{diagram}(c) by pathway \uppercase\expandafter{\romannumeral6}. At this stage, the ionic states of pathways \uppercase\expandafter{\romannumeral2} and \uppercase\expandafter{\romannumeral6}, as well as \uppercase\expandafter{\romannumeral3} and \uppercase\expandafter{\romannumeral5}, become identical, allowing for interference between these pathways. In contrast to the case involving only pathways \uppercase\expandafter{\romannumeral2} and \uppercase\expandafter{\romannumeral4}, the inclusion of this interference can result in an asymmetric angle- and energy-resolved photoelectron spectrum. Intrinsically, Rabi oscillations fundamentally transform the entanglement between the ion state and the photoelectron, leading to new interference effects between distinct ionization pathways. This opens the possibility for coherent control of the photoelectron spectrum by varying the relative phase and time delay between the $\omega$ and $2\omega$ pulses. 

\begin{figure*}[hbtp]
\centering
\includegraphics[width=\textwidth]{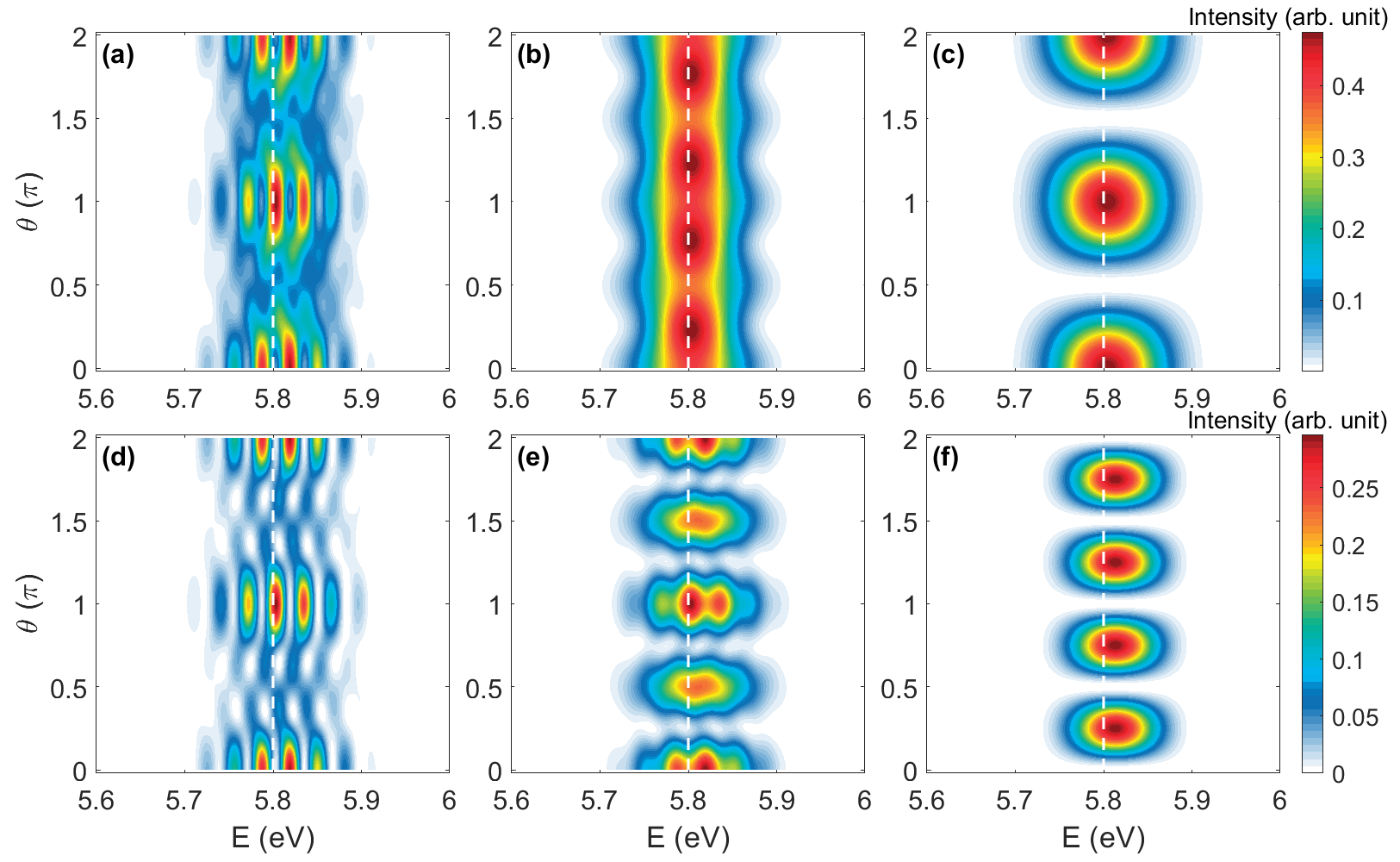}
\captionsetup{width=\textwidth}
\caption{(a) The PMD of a neon atom exposed to a $\omega+2\omega$ pulse pair with laser parameters $\omega=27.22\;\mathrm{eV}$, $I_{\omega}=8.9\times10^{11}\;\mathrm{W/cm^2}$, $I_{2\omega}=2.0\times10^{13}\;\mathrm{W/cm^2}$, $\delta=0$, and $\tau=120\;\mathrm{fs}$. (b) Same as (a), but with only the $\omega$ pulse. (c) Same as (a), but with only the $2\omega$ pulse. (d) The photoelectron spectrum resulting from the coherent superposition of pathways \uppercase\expandafter{\romannumeral2} and \uppercase\expandafter{\romannumeral6}. (e) The photoelectron spectrum resulting from the coherent superposition of pathways \uppercase\expandafter{\romannumeral3} and \uppercase\expandafter{\romannumeral5}. (f) The photoelectron spectrum resulting from the coherent superposition of pathways \uppercase\expandafter{\romannumeral1}. The white dashed line marks the photoelectron energy at 5.8 eV.}
\label{PMD}
\end{figure*}

In our calculations, $\omega$ is set to $27.22$ eV. The laser intensities for the fundamental ($I_{\omega}$) and second-harmonic ($I_{2\omega}$) pulses are $8.9\times10^{11}\;\mathrm{W/cm^2}$ and $2.0\times10^{13}\;\mathrm{W/cm^2}$, respectively. These intensities are chosen to ensure that the contributions from pathways \uppercase\expandafter{\romannumeral2} and \uppercase\expandafter{\romannumeral6} are comparable, thereby resulting in noticeable interference. Both pulses have sin-squared intensity profiles, with the $\omega$ pulse having a full width at half maximum (FWHM) of approximately 30 fs and the $2\omega$ pulse having an FWHM of around 45 fs.

Figure~\ref{PMD}(a) shows the PMD induced by the $\omega+2\omega$ pulse pair near $E_0$, with the relative phase and time delay between the two pulses set to 0 and -120 fs (the negative sign indicates that the $2\omega$ pulse precedes the $\omega$ pulse), where $\theta$ represents the photoelectron emission angle. Two prominent features are observed in the photoelectron spectrum. First, it exhibits interference fringes in the photoelectron energy with a resolution of $\Delta E<0.1$ eV, as the large time delay between the laser pulses induces temporal double-slit interference. Second, the photoelectron spectrum displays significant left-right asymmetry, with strong differences between the photoelectron spectrum in the angular ranges $\theta \in [0,\pi/2]$ and $\theta \in (\pi/2,\pi]$. At the position of the white dashed line in Fig.~\ref{PMD}(a), a local minimum appears at $\theta=0$, while a local maximum occurs at $\theta=\pi$. Based on our analysis of the contributing channels, this asymmetry is due to interference between pathways \uppercase\expandafter{\romannumeral2} and \uppercase\expandafter{\romannumeral6} or between pathways \uppercase\expandafter{\romannumeral3} and \uppercase\expandafter{\romannumeral5}. To verify this, we plot the photoelectron spectra for the cases where either only an $\omega$ pulse or a $2\omega$ pulse is applied, as shown in Figs.~\ref{PMD}(b) and \ref{PMD}(c). For these two subfigures, the points at $\theta=0$ and $\theta=\pi$ on the white dashed line are symmetric with respect to $\theta=\pi/2$. No interference fringes or asymmetric patterns are observed in these cases, which thus confirms that these features result from the combined effect of an $\omega$ photon and a $2\omega$ photon.

We categorize the signals based on their ionic states: pathways \uppercase\expandafter{\romannumeral2} and \uppercase\expandafter{\romannumeral6} correspond to the ionic state $\ket{2p_0^{-1}}$, pathways \uppercase\expandafter{\romannumeral3} and \uppercase\expandafter{\romannumeral5} correspond to $\ket{2s_0^{-1}}$, and pathways \uppercase\expandafter{\romannumeral1} or \uppercase\expandafter{\romannumeral4} correspond to $\ket{2p_{\pm 1}^{-1}}$. The corresponding results are shown in Figs.~\ref{PMD}(d), \ref{PMD}(e), and \ref{PMD}(f). Except for Fig.~\ref{PMD}(f), both Figs.~\ref{PMD}(d) and \ref{PMD}(e) display the two features. The interference between pathways \uppercase\expandafter{\romannumeral2} and \uppercase\expandafter{\romannumeral6}, or between pathways \uppercase\expandafter{\romannumeral3} and \uppercase\expandafter{\romannumeral5}, requires contributions from both $\omega$ and $2\omega$ pulses, leading to double-slit interference in the time domain. Furthermore, these pathways involve partial wave interference between the $s_0+d_0$ and $p_0$ waves, leading to asymmetric behavior of the photoelectron spectra. The different interference patterns observed in Figs.~\ref{PMD}(d) and \ref{PMD}(e) reveal phase differences between the partial waves corresponding to the ionic states $\ket{2s_0^{-1}}$ and $\ket{2p_0^{-1}}$. However, we notice that the signal from pathways involving $\ket{2p_0^{-1}}$ dominates. This may be due to the duration of the $\omega$ pulse, which allows most of the $\ket{2s_0^{-1}}$ state to convert to $\ket{2p_0^{-1}}$. The incoherent summation of contributions from other pathways with different ionic states leads to the final photoelectron spectrum shown in Fig.~\ref{PMD}(a).

\begin{figure}[h]
\flushleft
\includegraphics[width=0.5\textwidth]{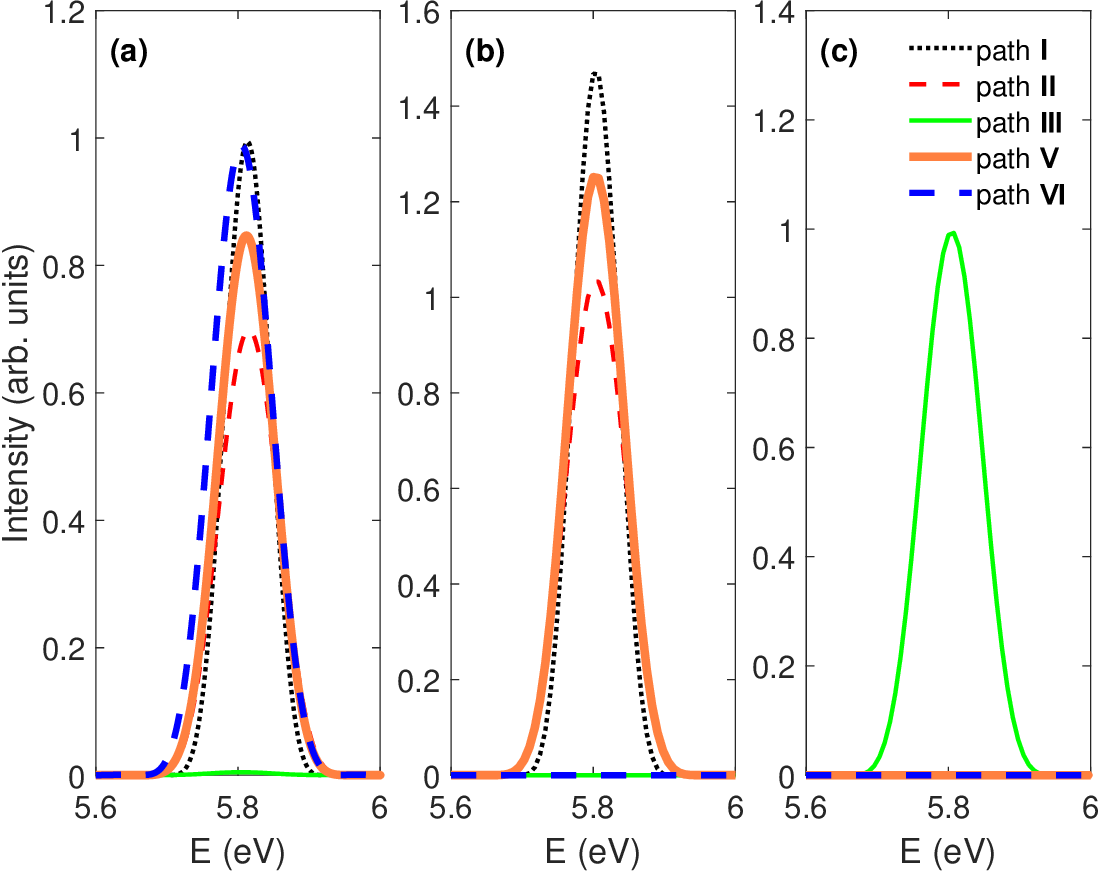}
\caption{(a) The partial wave intensity of different paths for $\omega=27.22\;\mathrm{eV}$, $I_{\omega}=8.9\times10^{11}\;\mathrm{W/cm^2}$, $I_{2\omega}=2.0\times10^{13}\;\mathrm{W/cm^2}$, $\delta=0$ and $\tau =120\;\mathrm{fs}$. The black dotted curve, red thin dashed curve, green thin solid curve, orange thick solid curve and, blue thick dashed curve represent pathways \uppercase\expandafter{\romannumeral1}, \uppercase\expandafter{\romannumeral2}, \uppercase\expandafter{\romannumeral3}, \uppercase\expandafter{\romannumeral5}, and \uppercase\expandafter{\romannumeral6}, respectively. (b) Same as (a), but with only the $\omega$ pulse. (c) Same as (a), but with only the $2\omega$ pulse.} 
\label{partial}
\end{figure}

To gain a deeper understanding of the photoelectron spectra, we present the partial wave amplitudes involved. Figures~\ref{partial}(a), \ref{partial}(b), and \ref{partial}(c) display photoelectron energy spectra (PES) for the cases of the $\omega+2\omega$ pulse pair, a single $\omega$ pulse, and a single $2\omega$ pulse, respectively. The contributions from different pathways are separated. The pathway \uppercase\expandafter{\romannumeral4} is omitted as it is identical to the pathway \uppercase\expandafter{\romannumeral1}. From Fig.~\ref{partial}(b), we observe that contributions from pathways \uppercase\expandafter{\romannumeral1}, \uppercase\expandafter{\romannumeral2}, \uppercase\expandafter{\romannumeral4}, and \uppercase\expandafter{\romannumeral5} are comparable. Although pathway \uppercase\expandafter{\romannumeral5} can be induced by the $\omega$ pulse alone, it does not interfere with other pathways due to differing ionic states. In the case of a single $2\omega$ pulse, only pathway \uppercase\expandafter{\romannumeral3} contributes. These results are consistent with our previous discussions. Interestingly, in the case of the $\omega+2\omega$ pulse pair, pathway \uppercase\expandafter{\romannumeral3} almost entirely converts to pathway \uppercase\expandafter{\romannumeral6}. The substantial contribution from pathway \uppercase\expandafter{\romannumeral6} allows for significant interference with the contribution from pathway \uppercase\expandafter{\romannumeral2}. The Rabi coupling induced by the $\omega$ pulse converts the entanglement into coherence, enabling us to coherently control the photoelectron spectra by varying the relative phases between the two pulses.

\begin{figure}[h]
\flushleft
\includegraphics[width=0.5\textwidth]{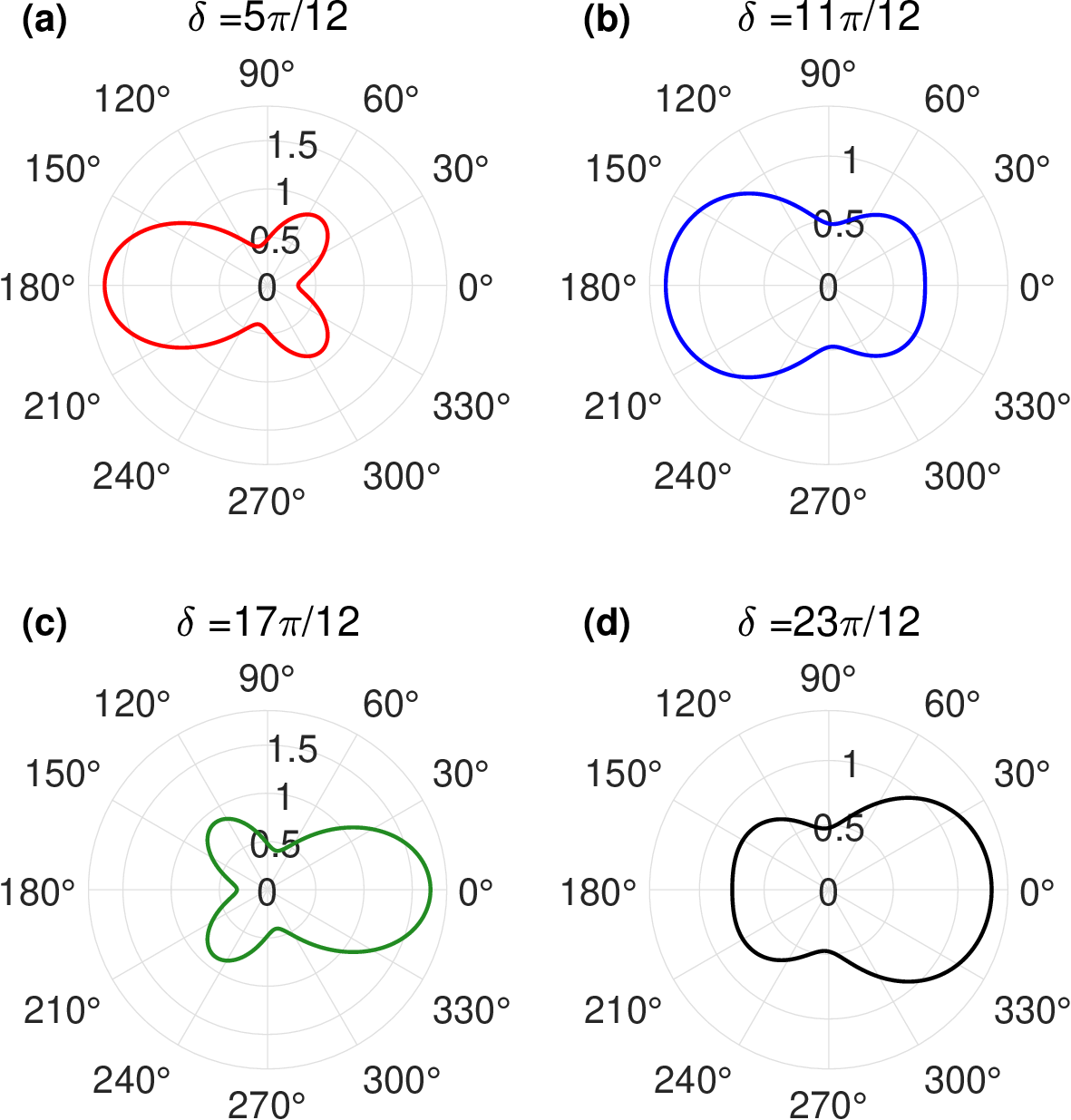}
\caption{The PADs at $E_{0}$, for different values of $\delta$: (a) $\delta=5\pi/12$; (b) $\delta=11\pi/12$; (c) $\delta=17\pi/12$; (d) $\delta=23\pi/12$.} 
\label{PADs}
\end{figure}

Then, we fix the time delay between the two pulses at 120 fs and vary the phase difference $\delta$ from $0$ to $2\pi$. Figure~\ref{PADs} presents four representative PADs at $E_0$ for $\delta=5\pi/12$, $11\pi/12$, $17\pi/12$, and $23\pi/12$, respectively. Distinct patterns of left-right asymmetry can be clearly observed, thus realizing the coherent control of the PAD by adjusting the relative phase between the two pulses.

\begin{figure}[h]
\flushleft
\includegraphics[width=0.47\textwidth]{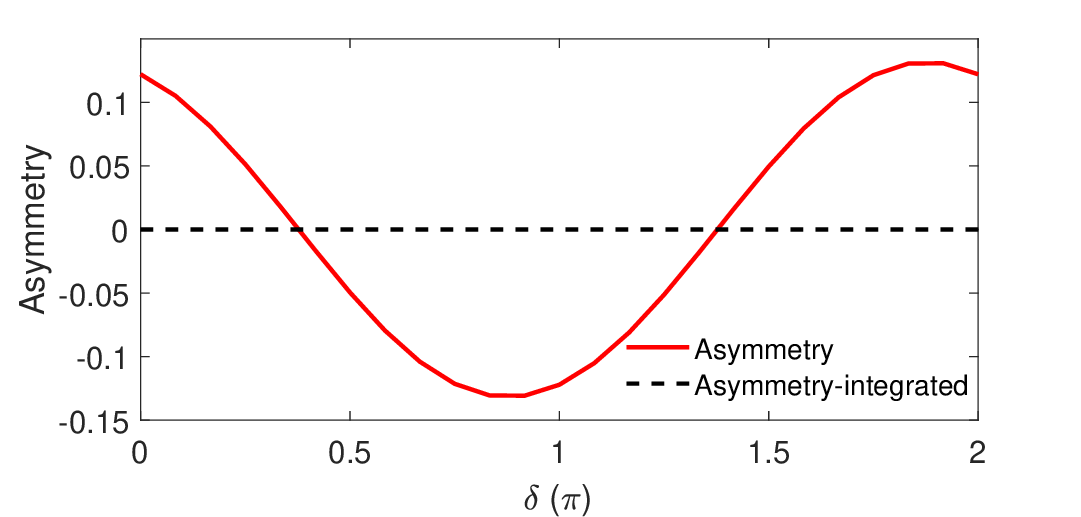}
\caption{The red solid curve represents the asymmetry of the energy-resolved electron emission for $\omega =27.22\;\mathrm{eV}$, $I_{\omega}=8.9\times 10^{11}\;\mathrm{W/cm^2}$, $I_{2\omega}=2.0\times10^{13}\;\mathrm{W/cm^2}$, $E_{0}=5.81\;\mathrm{eV}$, and $\tau=120\;\mathrm{fs}$. It is a function of the relative phase $\delta$. The black dashed line represents the integral of the asymmetry for all energies near $E_{0}$.} 
\label{Asymmetry}
\end{figure}

To more accurately quantify the controlled asymmetry in Figs.~\ref{PMD} and Figs.~\ref{PADs}, the asymmetry $A(\delta, E)$ is defined as the difference in the PAD at photoelectron energy $E$ between the two hemispheres ($0<\theta<\pi/2$ and $\pi/2<\theta<\pi$) emissions divided by their summation~\cite{ishikawa2023control}, namely
\begin{equation}
    A(\delta, E) = \frac{\int_{0}^{\frac{\pi}{2}}I(\theta, E)d\theta-\int_{\frac{\pi}{2}}^{\pi}I(\theta, E)d\theta}{\int_{0}^{\frac{\pi}{2}}I(\theta, E)d\theta+\int_{\frac{\pi}{2}}^{\pi}I(\theta, E)d\theta},
    \label{Asymmetry01}
\end{equation}
where $I(\theta, E)$ is the PAD at photoelectron energy $E$. The asymmetry can also be expressed in terms of the $\beta$ parameters as,
\begin{equation}
    A(\delta,E) = \frac{\beta_{1}}{2}- \frac{\beta_{3}}{8}.
    \label{Asymmetry02}
\end{equation}
The $\beta$ parameters can be extracted from calculations with different $\delta$ by projecting the PADs onto the Legendre polynomials~\cite{you2020new,mao2024decoupling,di2019complete}, given by $\beta_k(E)=\int_0^{\pi}I(\theta,E)P_k(\cos\theta)d\theta~(k=1,3)$, where $P_k(\cos\theta)$ is the Legendre polynomial of order $k$. The asymmetry at $E_0$, denoted as $A(\delta, E_0)$, as a function of $\delta$, is shown by the red solid line in Fig.~\ref{Asymmetry}. It is observed that the minimum value occurs when $\delta$ is between $2.62$ and $2.88$, which is consistent with the results presented in Ref.~\cite{ishikawa2023control}. Figures~\ref{PADs}(a) and \ref{PADs}(c) correspond to the PADs when the asymmetry approaches zero, while Figs.~\ref{PADs}(b) and \ref{PADs}(d) show the PADs at the minimum and maximum of $A(\delta, E_0)$, respectively. Interestingly, the PADs remain asymmetric even when the total asymmetry is close to zero (e.g., at $\delta = 5\pi/12$ and $\delta = 17\pi/12$). This phenomenon likely arises from the fact that the total asymmetry only characterizes the difference in photoelectron emission between the two hemispheres and does not fully reflect the entire structure of the PAD.

In the energy range near $E_0$, the PADs do not exhibit the same asymmetry as at $E_0$ because the amplitudes and phases of different partial waves vary with the photoelectron energy. Notably, if the asymmetries corresponding to these different energies are integrated, the final result becomes symmetric. This is demonstrated by the black dashed line in Fig.~\ref{Asymmetry}, which represents the integration of the PADs in the photoelectron energy range from 5.6 eV to 6.0 eV. Regardless of changes in $\delta$, the result remains zero.

\begin{figure*}[hbtp]
\centering
\includegraphics[width=\textwidth]{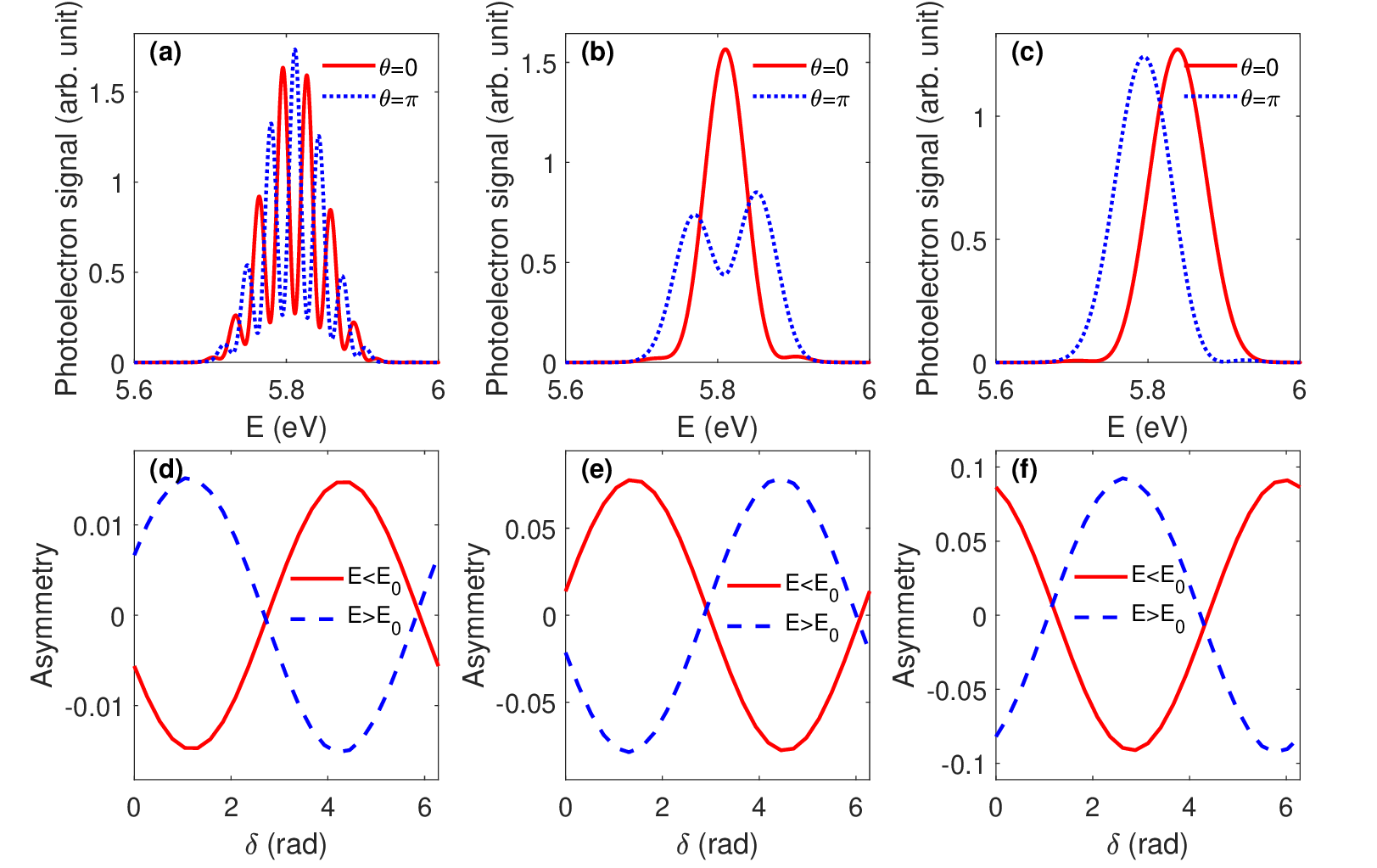}
\captionsetup{width=\textwidth}
\caption{(a) Photoelectron signal for $\omega=27.22\;\mathrm{eV}$, $I_{\omega}=8.9\times10^{11}\;\mathrm{W/cm^2}$, $I_{2\omega}=2.0\times10^{13}\;\mathrm{W/cm^2}$, $\delta=0$ and $\tau =120\;\mathrm{fs}$. The red solid curve represents $\theta=0$, and the blue dashed curve represents $\theta = \pi$; (b) Same as (a), but with a time delay of $\tau=20\;\mathrm{fs}$. (c) Same as (a), but with a time delay of $\tau=0\;\mathrm{fs}$. (d) The integration of asymmetry over different energy regions at a time delay of 120 fs. The blue dashed line represents the integral of the asymmetry for energies greater than the central energy $E_0$, while the red solid line represents the integral of the asymmetry for energies less than $E_0$. (e) Same as (d), but with a time delay of $\tau=20\;\mathrm{fs}$. (f) Same as (d), but with a time delay of $\tau=0\;\mathrm{fs}$.} 
\label{Signal}
\end{figure*}

Revisiting the earlier analysis, we find that the interference fringes in the photoelectron energy domain originate from double-slit interference in the time domain, which results from the combined effect of a $2\omega$ photon and a delayed $\omega$ photon. To validate this, we calculate the photoionization induced by the $\omega+2\omega$ pulse, similar to the previous case, but with pulse delays of $\tau = 20$ fs and $\tau = 0$ fs. The photoelectron signals at $\theta = 0$ and $\pi$ for delays of $\tau = 120$ fs, $20$ fs, and $0$ fs are shown in Figs.~\ref{Signal}(a), \ref{Signal}(b), and \ref{Signal}(c), respectively. The red solid curve represents the $\theta = 0$ signal, while the blue dashed curve corresponds to the $\theta=\pi$ signal. We can clearly observe that the fringe structure weakens as the pulse delay decreases, and when the delay is zero, the fringes completely vanish.

The oscillatory fringes in the photoelectron energy domain require very high energy resolution ($\Delta \varepsilon<0.1$ eV), which is difficult to achieve in real experiments. If the resolution is not high enough, the detected photoelectron signal becomes an integrated result over a small energy range, which can average the asymmetric behavior, as demonstrated in Fig.~\ref{Asymmetry}. To make the asymmetry of our coherent control more clearly observable, we propose using the half-integrated asymmetry. Instead of integrating the PADs over the full energy range near $E_0$, we split the range into two subgroups: one for $E<E_0$ and the other for $E>E_0$. Figures~\ref{Signal}(d), \ref{Signal}(e), and \ref{Signal}(f) show how these two values depend on $\delta$ for delays of 120 fs, 20 fs, and 0 fs, respectively. The results vary with delay, as the delay also affects the relative phases between the two pulses. We can find that the signature of the controlled asymmetry remains after this operation.

In real experiments, for a delay of $\tau=120$ fs, part of the energy-resolved information may be lost due to limited resolution. Since the real structure oscillates rapidly as a function of photoelectron energy, this loss of information leads to the energy-integrated results being nonphysical. However, as shown in Fig.~\ref{Signal}, for $\tau=0$ fs, even though no clear signatures of the PADs are visible, coherent control of the PADs is still maintained through the half-integrated asymmetry. As the photoelectron signals at $\theta=0$ and $\pi$ change smoothly with the photoelectron energy, the physical results of the half-integrated asymmetry can be preserved, even if the resolution is insufficient. In conclusion, this coherent control can be experimentally observed by choosing the $\omega+2\omega$ pulse with zero time delay and using the half-integrated asymmetry as the observable quantity.

\section{Conclusion \label{secIV}}

We have performed an \textit{ab initio} calculation of the ionization process of a neon atom under the influence of a $\omega+2\omega$ pulse pair, where $\omega$ was chosen to be the resonant energy between the $2s$ and $2p$ states of the neon atom. Our focus was on the photoelectrons emitted from these two subshells with the same energy (around $E_{0}=\omega-I_{p}$), and we found that the photoelectron spectra exhibited interference patterns when Rabi coupling was taken into account. The fundamental frequency $\omega$ induces Rabi coupling between these two subshells. If the Rabi coupling is neglected, no interference will occur between the photoelectrons from different pathways, because the ionic state and the photoelectron are entangled. Due to the Rabi coupling, some different ionic states merge into the same ionic state, leading to interference between them. This conclusion is also evident from the photoelectron spectra analysis and provides the possibility of achieving coherent control by varying the relative phase of the $\omega+2\omega$ pulse pair.

By fixing the pulse delay between the two pulses and varying the phase difference between them, we observe that the PAD shows asymmetry between the left and right sides. The observed asymmetry clearly demonstrates that the Rabi coupling transforms the entanglement into a coherent superposition, thereby allowing for coherent control of the photoelectron angular distribution by adjusting the relative phase between the $\omega$ and $2\omega$ pulses.

We believe that this work can inspire experimental investigations using FELs to study ion-photoelectron coherence and entanglement. Therefore, a brief discussion of the feasibility of such an experiment is essential. Firstly, the laser parameters used in our simulations are achievable with modern FELs. However, conducting a complete experiment requires extremely high energy resolution, which is difficult to achieve in real experiments. To mitigate this challenge, we propose using the asymmetry integrated over a limited energy range, rather than integrating over the full energy range. We have verified that by separately integrating the asymmetry for energies below and above the central energy, the half-integrated asymmetry still reveals the asymmetric behavior. This approach makes the observation of PAD asymmetry in experiments more feasible.

\begin{acknowledgments}
This work was supported by National Natural Science Foundation of China (NSFC) (Grant No. 11925405, 12274294). 
This research was also supported in part by a Grant-in-Aid for Scientific Research (Grant No. JP19H00869, JP20H05670, JP22H05025, JP22K18982, and JP24H00427) from the Ministry of Education, Culture, Sports, Science and Technology (MEXT) of Japan. This research was also partially supported by the MEXT
Quantum Leap Flagship Program, (Grant No. JPMXS0118067246)
and JST COI-NEXT, (Grant No. JPMJPF2221).
The computations in this paper were run on the Siyuan-1 cluster supported by the Center for High Performance Computing at Shanghai Jiao Tong University.
\end{acknowledgments}

\section*{Data availability}
The data that support the findings of this work is openly available \cite{data}.

\bibliography{ref}

\end{document}